\documentclass[preprint,showpacs,twocolumn]{revtex4}
\usepackage{graphicx}
\usepackage{amsmath}
\usepackage{amssymb}
\usepackage{bm}
\newcommand{\cH}{\mathcal{H}}
\newcommand{\R}{\mathbb{R}}
\newcommand{\vb}{\bm{b}}
\newcommand{\vk}{\bm{k}}
\newcommand{\vq}{\bm{q}}
\newcommand{\vx}{\bm{x}}
\newcommand{\vy}{\bm{y}}
\def\bra{\langle}
\def\ket{\rangle}
\def\d{\partial}
\def\dk#1#2{\frac{ d^{#2}{#1} }{ (2\pi)^{#2} }} 
\begin{document}
\title{On quantization in light-cone variables compatible with wavelet transform}
\author{M.V.Altaisky}
\affiliation{Space Research Institute RAS, Profsoyuznaya 84/32, Moscow, 117997, Russia}
\email{altaisky@mx.iki.rssi.ru}
\author{N.E.Kaputkina}
\affiliation{National Technological University ''MISIS'', Leninsky prospect 4, Moscow, 119049, Russia}
\email{nataly@misis.ru}

\date{Sep 3, 2015}

\begin{abstract}
Canonical quantization of quantum field theory models is inherently 
related to the Lorentz invariant partition  of classical fields into the positive and the negative frequency parts 
$
u(x) = u^+(x) + u^-(x),
$
performed with the help of Fourier transform in Minkowski space. 
That is the commutation relations are being established between non 
localized solutions of field equations. At the same time the construction 
of divergence free physical theory requires the separation of the contributions 
of different space-time scales. In present paper, using the light-cone variables, we propose a quantization procedure 
which is compatible with separation of scales using continuous wavelet 
transform, as described in our previous paper \citep{AK2013}. 
\end{abstract}

\maketitle

\section{Introduction}
The construction of quantum field theory models is inherently related to the Lorentz invariant partition  of classical fields into the positive and the negative frequency parts 
$$
u(x) = u^+(x) + u^-(x), \quad x \in \R^{1,3}
$$
The quantization procedure itself is then based on the commutators of the energy-momentum 
tensor components $T^{0\mu}$ with the positive and the negative frequency field operators
$u^\pm(\vk), \vk \in \R^3$, with the field equations being used to eliminate the redundant 
component of the quantum fields.

The calculation of the $n$-point Green functions  $\bra u(x_1) \ldots u(x_n) \ket$, 
the functional derivatives of the generating functional, is well known to suffer from loop divergences in both UV and IR domains of momentum space. The way to rule out the divergences is to 
separate the contributions of different scales, which can be formally casted  in the form 
\cite{Altaisky2010PRD}
$$
u(x) = \int u_a(x) \frac{da}{a},
$$
where the ''scale component'' $u_a(x)$ is not yet well defined. The most known way to separate 
the scales is the renormalization group technique \cite{GL1954,Wilson1973} the less known is the wavelet transform in quantum 
field theory \cite{AK2013}. 

The consideration would be straightforward for Euclidean quantum field theory, where the projection 
of an arbitrary function $u(x) \in L^2(\R^d)$ onto the scale $a$ is given by the convolution 
\begin{equation}
u_a(b) := \int_{\R^d} \frac{1}{a^d} \bar{g} \left(\frac{x-b}{a}\right) u(x) d^d x,
\end{equation} 
so that the function $u(x)$ can be reconstructed from the set of its {\sl wavelet coefficients}
$\{ u_a(b)\}$ by 
the {\sl inverse wavelet transform} \cite{Daub10}
\begin{equation}
u(x) = \frac{1}{C_g} \int_{\R_+ \times \R^d} \frac{1}{a^d} g\left(\frac{x-b}{a}\right) u_{a}(b) \frac{dad^db}{a}  \label{iwt}
\end{equation}
The analyzing function $g(x)$,  satisfying  
 rather loose admissibility condition 
$
\int \frac{|\tilde{g}(ak)|^2}{a} da = C_g < \infty
$
is usually referred to as a basic wavelet.

The continuous wavelet transform (CWT) is a feasible alternative to the usual Fourier 
transform 
\begin{equation}
u(x) = \int e^{\imath k x} \tilde{u}(k) \frac{d^dk}{(2\pi)^d}
\label{FT}
\end{equation} 
because in any measurement are not accessible exactly in a given point $x$: 
to localize a particle in an interval $\Delta x$ the measuring device 
requests a momentum transfer of order $\Delta p\!\sim\!\hbar/\Delta x$. If $\Delta x$ is too small the field $u(x)$ at 
a fixed point $x$ has no experimentally verifiable meaning. 
At the same time establishing of canonical commutation relations between the field operators 
is essentially based on Fourier transform \eqref{FT}.
It is intuitively clear that the commutator $[u_{a_1}(b_1),u_{a_2}(b_2)]$ is a function 
of $\frac{a_1}{a_2}$, which vanish if $\log \frac{a_1}{a_2}$ is significantly different from zero. This fact is well known in radiophysics : if a field (system) is localized in a region of size $a_1$ centered at point $b_1$, it may be detected by other field with significant 
probability only in case when $a_1$ and $a_2$ have the same order. If the window width $a_2$ 
is too narrow or too wide in comparison to $a_1$ the probability of detection is low. 

In the remainder of this paper we present the derivation of the canonical commutation 
relations between the field operators describing massive scalar field that depend on both 
position and resolution in $\R^{1,3}$ Minkowski space.

\section{Continuous wavelet transform}
\subsection{Basics of the continuous wavelet transform}
Let $\cH$ be a Hilbert space of states for a quantum field $|\phi\ket$. 
Let $G$ be a locally compact Lie group acting transitively on $\cH$, 
with $d\mu(\nu),\nu\in G$ being a left-invariant measure on $G$. Then, 
similarly to representation of a vector $|\phi\ket$ in a Hilbert space 
of states $\cH$ as a linear combination of an eigenvectors of momentum 
operator 
$
|\phi\ket=\int |p\ket dp \bra p |\phi\ket,$
any $|\phi\ket \in \cH$ can be decomposed with respect to 
a representation $U(\nu)$ of $G$ in $\cH$ \cite{Carey1976,DM1976}:
\begin{equation}
|\phi\ket= \frac{1}{C_g}\int_G U(\nu)|g\ket d\mu(\nu)\bra g|U^*(\nu)|\phi\ket, \label{gwl} 
\end{equation} 
where $|g\ket \in \cH$ is referred to as an admissible vector, 
or {\em basic wavelet}, satisfying the admissibility condition 
$$
C_g = \frac{1}{\| g \|^2} \int_G |\bra g| U(\nu)|g \ket |^2 
d\mu(\nu)
<\infty. 
$$
The coefficients $\bra g|U^*(\nu)|\phi\ket$ are referred to as 
wavelet coefficients. 

If the group $G$ is abelian, the wavelet transform \eqref{gwl} with 
$G:x'=x+b'$ coincides with Fourier transform. 

\subsection{Euclidean space}
The next to the abelian group is the group of the affine transformations 
of the Euclidean space $\R^d$
\begin{align}\nonumber 
G: x' = a R(\theta)x + b, \\ 
x,b \in \R^d, a \in \R_+, \theta \in SO(d), \label{ag1}
\end{align} 
where $R(\theta)$ is the rotation matrix.
We define unitary representation of the affine transform \eqref{ag1} with 
respect to the basic wavelet $g(x)$ as follows:
\begin{equation}
U(a,b,\theta) g(x) = \frac{1}{a^d} g \left(R^{-1}(\theta)\frac{x-b}{a} \right).
\end{equation}  
(We use $L^1$ norm \cite{Chui1992,HM1996} instead of usual $L^2$ to keep the physical dimension 
of wavelet coefficients equal to the dimension of the original fields).

Thus the wavelet coefficients of the function $u(x) \in L^2(\R^d)$ with 
respect to the basic wavelet $g(x)$ in Euclidean space $\R^d$ can be written 
as 
\begin{equation}
u_{a,\theta}(b) = \int_{\R^d} \frac{1}{a^d} \overline{g \left(R^{-1}(\theta)\frac{x-b}{a} \right) }u(x) d^dx. \label{dwtrd}
\end{equation} 
The wavelet coefficients \eqref{dwtrd} represent the result of the measurement 
of function $u(x)$ at the point $b$ at the scale $a$ with an aperture 
function $g$ rotated by the angle(s) $\theta$ \cite{PhysRevLett.64.745}.

The function $u(x)$ can be reconstructed from its wavelet coefficients 
\eqref{dwtrd} using the formula \eqref{gwl}:
\begin{widetext}
\begin{equation}
u(x) = \frac{1}{C_g} \int \frac{1}{a^d} g\left(R^{-1}(\theta)\frac{x-b}{a}\right) u_{a\theta}(b) \frac{dad^db}{a} d\mu(\theta) \label{iwt}
\end{equation}
\end{widetext}
The normalization 
constant
$C_g$ is readily evaluated using Fourier transform:
\begin{align*}
C_g &=& \int_0^\infty |\tilde g(aR^{-1}(\theta)k)|^2\frac{da}{a} d\mu(\theta)\\ 
&=& \int |\tilde g(k)|^2 \frac{d^dk}{|k|^d}<\infty.
\end{align*}
For isotropic wavelets 
$$
C_g = \int_0^\infty |\tilde g(ak)|^2\frac{da}{a}
= \int |\tilde g(k)|^2 \frac{d^dk}{S_{d}|k|^d},
$$
where $S_d = \frac{2 \pi^{d/2}}{\Gamma(d/2)}$ is the area of unit sphere 
in $\R^d$.

It is helpful to rewrite continuous wavelet transform in Fourier form:
\begin{eqnarray*}
u(x) &=& \frac{1}{C_g} \int_0^\infty \frac{da}{a} \int \dk{k}{d} e^{\imath k x}
\tilde g(ak) \tilde u_a(k), 
\\  
\tilde u_a(k) &=& \overline{\tilde g(ak)}\tilde u(k) .
\end{eqnarray*}

The wavelet function $\tilde{g}(ak)$ works as a {\em band-pass filter}, which injects a part of the energy carried by the k-mode of the function $u(x)$ into the ''detector'' of scale $a$, depending on how the product $|ak|$ is different from the unity. 

Indeed, taking the plane wave $\phi(x) = (2\pi)^{-d} exp(\imath k_0 x)$ as 
an example of free particle with momentum $k_0$, so that $\hat{P} \phi(x) = k_0 \phi(x), 
\hat{P} = -\imath \d_x$, we get 
$$
\tilde{\phi}(k) = \delta^d(k-k_0), \quad \tilde{\phi}_a(k) = \bar{\tilde{g}}(ak) \delta^d(k-k_0)
$$  
and hence 
\begin{equation}
\phi_a(b) = e^{\imath k_0 b} \bar{\tilde{g}}(ak_0)
\end{equation}
The partial momentum per octave is $k_0 \frac{|\tilde g(ak_0)|^2}{C_g}$, so that the sum over all 
possible scales is $k_0$. 

It is impossible however to do such separation in Minkowski space $\R^{1,3}$ in space-time coordinates $(t,x,y,z)$. 

\subsection{Minkowski space}
To construct wavelet transform in Minkowski space it is convenient to turn  
from the space-time coordinates $x^\mu=(t,x,y,z)$ to the {\em light-cone coordinates}:
\begin{equation}x^\mu=(x_+,x_-,y,z), 
x_\pm = \frac{t \pm x}{\sqrt{2}},\vx_\perp = (y,z).
\label{lcc}
\end{equation}
This is the so-called infinite momentum frame. 
The advantage of the coordinates \eqref{lcc} for the calculations
in quantum field theory is significant simplification of the vacuum structure \cite{ChangMa1969,KS1970}.
The metrics in the light-cone coordinates becomes 
$$
g_{\mu\nu} = \begin{pmatrix}
0 & 1 & 0 & 0 \cr 
1 & 0 & 0 & 0 \cr 
0 & 0 &-1 & 0 \cr
0 & 0 & 0 & -1
\end{pmatrix}.
$$
The rotation matrix -- the Loretnz boosts in $x$ direction and the rotations in $(y,z)$ plane -- has a block-diagonal form 
$$
M(\eta,\phi) = \begin{pmatrix}
e^{\eta} & 0 & 0 & 0 \cr
0 & e^{-\eta} & 0 & 0 \cr
0 & 0 & \cos\phi & \sin\phi \cr
0 & 0 &-\sin\phi & \cos\phi
\end{pmatrix}, 
$$
so that $M^{-1}(\eta,\phi)=M(-\eta,-\phi)$. Hyperbolic rotation in ($t,x$) plane 
is determined by the hyperbolic rotation angle -- the rapidity $\eta$. 
The rotations in the transverse plane, not affected by the Lorentz 
contraction,  are determined by the rotation angle $\phi$.

The Poincare group can be extended by the scale transformations $x'=a x$ 
to the affine group 
$$
x' = a M(\eta,\phi) x + b, $$
with the representation written in the same form as that of 
wavelet transform in Euclidean space $\R^d$, viz:
$$
U(a,b,\eta,\phi)u(x) = \frac{1}{a^4}u\left(M^{-1}(\eta,\phi)\frac{x-b}{a} \right), 
$$
defined in $L^1$ norm in accordance to \cite{HM1996,AK2013}.
So, we have straightforward generalization of the definition of wavelet 
coefficients of a function $f(x)\in L^2(\R^{1,3})$ 
with respect to the basic wavelet $g$ \cite{AK2013iv,AK2013}
\begin{widetext}
\begin{align}
W_{a,b,\eta,\phi}[f] = \int dx_+ dx_- d^2 \vx_\perp 
\frac{1}{a^4} \overline{ g\left(M^{-1}(\eta,\phi)\frac{x-b}{a} \right)} f(x_+,x_-,\vx_\perp). 
\label{dwtm}
\end{align}
\end{widetext}

The difference from calculations in Euclidean space $\R^4$ is 
that the basic wavelet $g(\cdot)$ cannot be defined globally on  
$\R^{1,3}$. Instead, it should be defined in four separate domains 
impassible by Lorentz rotations:
\begin{align*} 
A_1:  k_+ >0, k_-<0; & 
A_2:  k_+ <0, k_- >0;\\
A_3:  k_+ >0, k_- >0;& 
A_4:  k_+ <0, k_-<0 ,
\end{align*}
where $k$ is wave vector, $k_\pm = \frac{\omega \pm k_x}{\sqrt{2}}$. 
Four separate wavelets should be defined in these four domains \cite{PG2011,PG2012}:
\begin{equation}
g_i(x) = \int_{A_i} e^{\imath k x} \tilde{g}(k) \dk{k}{4}, \quad i=\overline{1,4}.
\end{equation}
We assume  the following definition of the Fourier transform 
in light cone coordinates:  
\begin{align*}\nonumber 
f(x_+,x_-,\vx_\perp) &=& \int e^{\imath k_- x_+ + \imath k_+ x_- -\imath \vk_\perp \vx_\perp}  \times \\ 
&\times& \tilde{f} (k_-, k_+,\vk_\perp)\frac{dk_+dk_-d^2\vk_\perp}{(2\pi)^4}. 
\end{align*}
Substituting the Fourier images into the definition \eqref{dwtm} 
we get 
\begin{align}\nonumber 
W_{ab\eta\phi}^i = \int_{A_i} e^{\imath k_- b_+ + \imath k_+ b_- -\imath \vk_\perp \vb_\perp} \tilde{f} (k_-, k_+,\vk_\perp) \\ \overline{\tilde{g}}(a e^\eta k_-, a e^{-\eta} k_+, a R^{-1}(\phi) \vk_\perp) \frac{dk_+dk_-d^2\vk_\perp}{(2\pi)^4}.
\end{align}
Similarly to the $\R^d$ case, the reconstruction formula is \cite{AK2013iv}:
\begin{widetext} 
\begin{align*}\nonumber 
f(x) 
 &=&  
\sum_{i=1}^4 \frac{1}{C_{g_i}} \int_{-\infty}^\infty d\eta 
\int_0^{2\pi} d\phi
\int_0^\infty \frac{da}{a} \int_{M^4_1} db_+ db_- d^2\vb_\perp 
\frac{1}{a^4} g_i \left( M^{-1}(\eta)\frac{\xi-b}{a} \right) W^i_{ab\eta\phi} \\
&=& \sum_{i=1}^4 \frac{1}{C_{g_i}} \int_{-\infty}^\infty d\eta  
\int_0^{2\pi} d\phi 
\int_0^\infty \frac{da}{a}
 \int_{A_i} \frac{dk_+dk_-d^2\vk_\perp}{(2\pi)^4}
e^{\imath k_- x_+ + \imath k_+ x_- -\imath \vk_\perp \vx_\perp} \times \\
&\times& 
\tilde{W}_{a\eta\phi}(k) \tilde{g}(ak_- e^\eta, a k_+ e^{-\eta},aR^{-1}(\phi)\vk_\perp)  
\end{align*}
\end{widetext}
\section{Quantization}
Same as in standard quantum field theory in Minkowski space we ought to 
use the mass-shell delta function to get rid of redundant degrees of freedom \cite{Bsh1980}. 
Let us consider the massive scalar field in $\R^{1,3}$ Minkowski space 
\begin{equation}
u(x) = \int e^{\imath k x} 2 \pi 
\delta(k^2-m^2) \tilde{u}(k_-,k_+,\vk_\perp) \frac{d^4k}{(2\pi)^4}
\label{uft}
\end{equation} 
The Lorentz invariant scalar product and the invariant volume in $k$-space
are  
\begin{align*}
kx \equiv k_0 x_0 - \vk\vx = k_- x_+ + k_+x_-
 - \vk_\perp \vx_\perp \\ 
\frac{d^4k}{(2\pi)^4} = \frac{dk_0 dk_x dk_y dk_z}{(2\pi)^4} = 
\frac{dk_-dk_+ d^2\vk_\perp}{(2\pi)^4}.
\end{align*}
For a massive scalar field because of the mass shell delta function 
$\delta(2k_+k_- -\vk^2 - m^2)$ only two domains $A_3$ and $A_4$ for which 
$k_+k_-$ is positive will contribute to the decomposition of $u(x)$.
The integration over the $k_-$ variable with the mass shell delta function 
gives 
$$
k_- = \frac{\vk_\perp^2+m^2}{2k_+}
$$
After the substitution of integration variable $k\to-k$ in integration 
over $A_4$, the decomposition of $u(x)$ takes the form 
\begin{align} \nonumber 
u(x) &=& \int \Huge[ e^{\imath k x}  
\tilde{u}\left(
\frac{\vk_\perp^2+m^2}{2k_+},k_+,\vk_\perp
\right) + \\ \nonumber 
&+& e^{-\imath k x} 
\tilde{u}\left(
-\frac{\vk_\perp^2+m^2}{2k_+},-k_+,-\vk_\perp
\right)\Huge] \times \\ \nonumber 
&\times& \theta(k_+)\frac{dk_+ d^2\vk_\perp}{2k_+(2\pi)^3} \\ \nonumber
&\equiv& \int \left[e^{\imath k x} \tilde u^+(k) + e^{-\imath k x} \tilde u^-(k) \right]
\times \\
&\times& \theta(k_+) \frac{dk_+ d^2\vk_\perp}{2 k_+(2\pi)^3} \label{pnd}
\end{align}
Both $\tilde u^+(k)$ and $\tilde u^-(k)$ are defined on a hemisphere in $\R^3$ and can be decomposed into scale components by continuous wavelet transform in Euclidean space.
The straightforward way to quantize the fields in the light-cone representation is to use the formal analogy between the decomposition 
\eqref{pnd} and the positive/negative frequency decomposition in usual 
coordinates ($t,\vx$) in the equal-time quantization scheme 
\begin{equation}\left.
\left\{u(t,\vx), \frac{\d L}{\d \dot{u}(t,\vy)}
\right\}\right|_{t=0} = \imath \delta^3 (\vx-\vy),
\label{pb}
\end{equation}
where the curly brackets stand for the Poisson brackets substituted 
by commutator (anti-commutator) for Bose (Fermi) quantum fields.

Using the Lagrangian 
\begin{equation}
L = \frac{\d u}{\d x_+} \frac{\d u}{\d x_-} - \frac{1}{2}(\d_\perp u)^2 - \frac{m^2}{2}u^2, \label{lpm}
\end{equation}
we can infer that the $x_+ = \frac{
t+x}{\sqrt{2}}$ variable can be considered as ''time'' on the light-cone 
\cite{LB1980}.
In analogy to common case the Poisson bracket can be then casted in the 
form
\begin{equation}
\left\{
u(x_+=0,x_-,x_\perp), \left. \frac{\d u}{\d y_-}\right|_{y_+=0}
\right\}=\imath \delta^3(x-y) \label{pblc}
\end{equation}
Substituting decomposition \eqref{pnd} into the bracket \eqref{pblc} and 
changing the bracket to commutator one gets 
\begin{equation}
\left[ \tilde u^-(k),\tilde u^+(q) \right] = 2k_+ (2\pi)^3 \delta^3(k-q). \label{cr1}
\end{equation}
The latter equation is different from the standard commutation relation 
by changing the energy ($k_0$) to the momentum ($k_+$). The role of energy is played by $k_-$ in the light-cone coordinates.

Substituting the inverse wavelet transform
\begin{equation}
\tilde u^\pm(k_+,\vk_\perp) = \frac{1}{C_g}\int_0^\infty \tilde g(ak) \tilde u^\pm_a(k) \frac{da}{a}, \label{fwt}
\end{equation} 
where $\tilde u^\pm_a(k) = \overline{\tilde g}(ak)\tilde u^\pm(k), k \equiv (k_+,\vk_\perp)$,
into the equality \eqref{cr1}, and assuming an isotropic basic wavelet $g(\cdot)$ for simplicity, we derive the commutation relations for the scale components
\begin{widetext}
\begin{equation}
\left[ \tilde u^-_{a_1}(k),\tilde u^+_{a_2}(q)
\right] = 16\pi^3 C_g a_1 \delta(a_1-a_2) k_+ \delta(k_+-q_+) \delta^2(\vk_\perp-\vq_\perp). \label{crs}
\end{equation}
\end{widetext}
The commutation relation \eqref{crs} meets the general form 
of  wavelet transform of the canonical commutation 
relations in Minkowski space, eq.(18) of \cite{AK2013}
\begin{align*}
[u^-_{ia\eta}(k),u^+_{ja'\eta'}(k') ] &=& a \delta(a-a') \delta (\eta-\eta')\times \\
&\times& \delta_{ij} C_{g_i} [u^-(k),u^+(k')],
\end{align*}
defined on four Lorentz-invariant domains $A_i,i=\overline{1,4}$. However, being defined on $k \in \R_+ \otimes \R^2$ it is easier for practical calculations. 

Introducing the vacuum state $\Phi_p$ with the momentum $p$ we get
\begin{align*}
P^n \tilde u^+(k) \Phi_p &=& (p^n+k^n) \tilde u^+(k) \Phi_p, \\
P^n \tilde u^-(k) \Phi_p &=& (p^n-k^n) \tilde u^-(k) \Phi_p.
\end{align*}
In the latter equations $\tilde u^\pm(k)$ can be subjected to 
wavelet transform so that $\tilde u^\pm(k)$ is expressed by \eqref{fwt}
with $k$ having only 3 independent components. In this way we can construct 
the {\em multiscale} Fock space of states 
\begin{widetext}
\begin{equation}  
\Phi = \sum_{j,s} \int F^{(\cdots j \cdots)}_s(a_1,k_1,\ldots,
a_s,k_s) 
\tilde u_{j_1 a_1}^+(k_1) \ldots \tilde u_{j_s a_s}^+(k_s) 
\frac{da_1dk_{1+} d^2\vk_{1\perp}}{a_1C_g16k_{1+}\pi^3} \ldots 
\frac{da_s dk_{s+} d^2\vk_{s\perp}}{a_sC_g16k_{s+}\pi^3} 
\Phi_0, 
\end{equation}
\end{widetext}
where $k_i = (k_{i+},k_{i\perp})$ are three dimensional vectors, $j$ denote 
all other indices of the quantum states, and 
$\Phi_0$ is a vacuum state $u^-_i(x) \Phi_0 = 0$. 

\section{Conclusions}
To be concluded, we have developed a quantization scheme suitable for applications in quantum theory of fields $u_a(x)$, which explicitly depend 
on both position $x$ and the scale (resolution) $a$. It is not suprising, 
that such fields can form a prospective framework for analytic calculations 
in quantum chromodynamics, where most approved results are obtained either 
numerically lattice simulations \cite{LQCD}, or analytically, with perturbation expansion being corrected by renormalization group methods \cite{CSS2004}. In the latter case the obtained results, viz., process amplitudes, parton distribution functions, 
nucleon form factors, tacitly depend on some formal scale parameter $\Lambda$, which is either cutoff momentum, or renormalization scale. 
From functional analysis point of view, this may suggest the use of space 
of functions which explicitly depend on both the position and the resolution. 
being operator-valued functions they certainly require commutation relations. 
The use of light-cone coordinates enables this construction. The massive scalar field quantization was choosen as a simple example. Perhaps the same 
technique can be used in general problems of quantum field theory, when 
wavelet transform is used to construct divergence free Green functions \cite{Federbush1995,Altaisky2010PRD,BP2013}.

\section*{Acknowledgement}
The work was supported in part by RFBR projects 13-07-00409, 14-02-00739 and by the Ministry of Education and Science of the Russian Federation  in the framework of Increase Competitiveness Program of MISiS.

\begin{thebibliography}{10}

\bibitem{AK2013}
M.~V. Altaisky and N.~E. Kaputkina.
\newblock Continuous wavelet transform in quantum field theory.
\newblock {\em Phys. Rev. D}, 88:025015, Jul 2013.

\bibitem{Altaisky2010PRD}
M.~V. Altaisky.
\newblock Quantum field theory without divergences.
\newblock {\em Phys. Rev. D}, 81:125003, 2010.

\bibitem{GL1954}
M.~Gell-Mann and F.~Low.
\newblock Quantum electrodynamics at small distances.
\newblock {\em Phys. Rev.}, 95:1300, 1954.

\bibitem{Wilson1973}
K.~G. Wilson.
\newblock Quantum field-theory models in less than 4 dimensions.
\newblock {\em Physical Review D}, 7(10):2911--2927, 1973.

\bibitem{Daub10}
I.~Daubechies.
\newblock {\em Ten lectures on wavelets}.
\newblock S.I.A.M., Philadelphie, 1992.

\bibitem{Carey1976}
A.~L. Carey.
\newblock Square-integrable representations of non-unimodular groups.
\newblock {\em Bull. Austr. Math. Soc.}, 15:1--12, 1976.

\bibitem{DM1976}
M.~Duflo and C.~C. Moore.
\newblock On regular representations of nonunimodular locally compact group.
\newblock {\em J. Func. Anal.}, 21:209--243, 1976.

\bibitem{Chui1992}
C.~K. Chui.
\newblock {\em An Introduction to Wavelets}.
\newblock Academic Press Inc., 1992.

\bibitem{HM1996}
C.~R. Handy and R.~Murenzi.
\newblock Continuous wavelet transform analysis of one-dimensional quantum
  bound states from first principles.
\newblock {\em Phys. Rev. A}, 54(5):3754--3763, 1996.

\bibitem{PhysRevLett.64.745}
E.~Freysz, B.~Pouligny, F.~Argoul, and A.~Arneodo.
\newblock Optical wavelet transform of fractal aggregates.
\newblock {\em Phys. Rev. Lett.}, 64(7):745--748, Feb 1990.

\bibitem{ChangMa1969}
Shau-Jin Chang and Shang-Keng Ma.
\newblock {Feynman Rules and Quantum Electrodynamics at Infinite Momentum}.
\newblock {\em Phys. Rev.}, 180:1506--1513, Apr 1969.

\bibitem{KS1970}
J.B. Kogut and D.E. Soper.
\newblock {Quantum electrodynamics in the infinite-momentum frame}.
\newblock {\em Phys. Rev. D}, 1(10):2901--2914, 1970.

\bibitem{AK2013iv}
M.V. Altaisky and N.E. Kaputkina.
\newblock On the wavelet decomposition in light cone variables.
\newblock {\em Russian Physics Journal}, 55(10):1177--1182, 2013.

\bibitem{PG2011}
E.A. Gorodnitskiy and M.V. Perel.
\newblock {The Poincare wavelet transform: Implementation and interpretation}.
\newblock In I.V. Andronov, editor, {\em {Proc. Int. Conf. Days on Difraction
  2011}}, pages 72--77, St.Petersburg, 2011. SPbU.

\bibitem{PG2012}
E.A. Gorodnitskiy and M.V. Perel.
\newblock Integral representations of solutions of the wave equation based on
  relativistic wavelets.
\newblock {\em J. Math. Phys.}, 45(38):385203, 2012.

\bibitem{Bsh1980}
N.~N. Bogoliubov and D.~V. Shirkov.
\newblock {\em Introduction to the theory of quantized fields}.
\newblock John Wiley, New York, 1980.

\bibitem{LB1980}
G.P. Lepage and S.J. Brodsky.
\newblock Exclusive processes in perturbative quantum chromodynamics.
\newblock {\em Phys. Rev. D}, 22(9):2157--2198, 1980.

\bibitem{LQCD}
T.~Degrand and C.~{DeTar}.
\newblock {\em Lattice Methods for Quantum Chromodynamics}.
\newblock World Scientific, 2006.

\bibitem{CSS2004}
J.~Collins, D.~Soper, and G.~Sterman.
\newblock Factorization of hard processes in {QCD}.
\newblock {\em Adv. Ser. Direct. High Energy Phys.}, 5:1--91, 1988.
\newblock arXiv.org:hep-ph/0409313.

\bibitem{Federbush1995}
P.~Federbush.
\newblock A new formulation and regularization of gauge theories using a
  non-linear wavelet expansion.
\newblock {\em Progr. Theor. Phys.}, 94:1135--1146, 1995.

\bibitem{BP2013}
F.~Bulut and W.N. Polyzou.
\newblock Wavelets in field theory.
\newblock {\em Phys. Rev. D}, 87:116011, 2013.

\end{thebibliography}

\end{document}